\documentclass[sort&compress,a4paper,oneside,3p,12pt]{melsarticle}

\usepackage{amssymb}
\usepackage{amsmath,amssymb,amsfonts,graphicx,float,multicol,fleqn}
\usepackage{multirow}

\newtheorem{lemma}{Lemma}

\begin{document}
\begin{frontmatter}

\title{Lagrangian method for solving Lane-Emden type equation arising in astrophysics on semi-infinite domains}


\author{K Parand}
\ead{k\_parand@sbu.ac.ir}
\author{A R Rezaei}
\ead{alireza.rz@gmail.com}
\author{A Taghavi\corref{cor}}
\cortext[cor]{Corresponding author. Tel:+98 21 22431653; Fax:+98 21 22431650.}
\ead{amirtaghavims2@yahoo.com}
\address{Department of Computer Sciences, Shahid Beheshti University, G.C., Tehran, Iran}

\begin{abstract}
In this paper we propose a Lagrangian method for solving Lane-Emden equation which
is a nonlinear ordinary differential equation on semi-infinite interval. This approach is based on a Modified generalized Laguerre functions Lagrangian method. The method reduces the solution of this problem to the solution of a system of algebraic equations.
We also present the comparison of this work with some well-known results and show that the present solution is acceptable.
\end{abstract}


\end{frontmatter}

\section{Introduction} 
Recently, spectral methods have been successfully applied in the approximation of differential boundary value problems defined in unbounded domains. We can apply different approaches using spectral methods to solve problems in semi-infinite domains.

The first approach is using Laguerre polynomials/functions \cite{Guo.num2000,Guo.Shen.Xu,Shen,Siyyam,Maday}.
Guo \cite{Guo.num2000} suggested a Laguerre-Galerkin method for the Burgers
equation and Benjamin-Bona-Mahony (BBM) equation on a
semi-infinite interval. It is shown that the Laguerre-Galerkin approximations are convergent on a semi-infinite interval with
spectral accuracy. He in \cite{Guo.Shen.Xu} introduced a new family of generalized Laguerre polynomials and investigated various orthogonal projections.
Shen \cite{Shen} proposed spectral methods using Laguerre functions and analyzed elliptic equations on regular unbounded domains. In \cite{Shen} is shown that spectral-Galerkin approximations based on Laguerre functions are stable and convergent with spectral accuracy in the Sobolev spaces. Siyyam \cite{Siyyam} applied two numerical methods for solving initial value problem differential equations using the Laguerre Tau method. Maday, et al. \cite{Maday} proposed a Laguerre type spectral method for solving partial differential equations.

The second approach is reformulating the original problem in a semi-infinite domain to a singular problem in a bounded domain by variable transformation and then using the Jacobi polynomials to approximate the resulting singular problem \cite{Guo.com2000}.

The third approach replacing the semi-infinite domain with $[0,K]$ interval by choosing $K$, sufficiently large. This method is named domain truncation \cite{Boyd2000}.

The fourth approach of spectral method is based on rational orthogonal functions. Boyd \cite{Boyd1987} defined a new spectral basis, named rational Chebyshev functions on the semi-infinite interval, by mapping to the Chebyshev polynomials. Guo et al. \cite{Guo.sci2000} introduced a new set of rational Legendre functions which is mutually orthogonal in $L^2(0,+\infty)$. They applied a spectral scheme using the rational Legendre functions for solving the Korteweg-de Vries equation on the half line. Boyd et al. \cite{Boyd2003} applied pseudospectral methods on a semi-infinite interval and compared rational Chebyshev, Laguerre, and mapped Fourier sine.

The authors of \cite{Parand.phy2004,Parand.mat2004,Parand.com2004} applied spectral method to solve nonlinear ordinary differential equations on semi-infinite intervals. Their approach was based on a rational Tau method. They obtained the operational matrices of derivative and product of rational Chebyshev, Legendre functions, then applied these matrices together with Tau method to reduce the solution of these problems to the solution of a system of algebraic equations. The authors of \cite{Parand.JCP} also applied pseudospectral method based on rational Legendre functions to solve Lane-Emden equations.

This paper is arranged as follows:\\
In section \ref{Lane-Emden.equation} we describe Lane-Emden equation.
In section \ref{Properties.of.MGLF} we describe the formulation of generalized Laguerre polynomials and modified generalized Laguerre functions required for our subsequent development.
In section \ref{Sec.Lagrangian.Interpolation} we detailed description of Lagrangian interpolants construction and properties then obtained the operational matrices of derivative of modified generalized Laguerre functions to applied these matrices together with the Lagrangian method to reduce the solution of this problem to the solution of the system of algebraic equation.
Section \ref{solving.Lane-Emden} summarizes the application of this method for solving Lane-Emden equation and a comparison is made with existing methods in the literature. The results show preference of this method in comparison with the others. The conclusions are described in the final section.
\section{Lane-Emden equation}\label{Lane-Emden.equation} 
In the study of stellar structure \cite{Chandrasekhar.Dover1957} an important
mathematical model described by the second-order ordinary differential equation
\begin{equation}\label{Eq.Lane-EmdenMain}
xy''+2y'+xg(y)=0,\qquad x>0,
\end{equation}
arises, where $g(y)$ is some given function of $y$.
Among the most popular form of $g(y)$ is
\begin{equation}\label{Eq.Lane-Emden}
g(y)=y^m,
\end{equation}
where $m$ is a constant. subject to the conditions
\begin{equation}\label{Eq.Lane-EmdenBoundry}
y(0)=1, \qquad y'(0)=0.
\end{equation}
This equation is standard Lane-Emden equation. It was first proposed by Lane \cite{Thomson} and studied in more detail by Emden \cite{Emden}.

The Lane-Emden equation describes a variety of phenomena in theoretical physics and astrophysics, including aspects of stellar structure,
the thermal history of a spherical cloud of gas, isothermal gas spheres, and thermionic currents \cite{Chandrasekhar.Dover1957}.

\subsection{Standard Lane-Emden equation}\label{Subsection.Lane-Emden equation}
This equation is one of the basic equations in the theory of stellar structure and has been the focus of many studies \cite{Bender.J. Math. Phys.1989,Biles,Bluman}.
This equation describes the temperature variation of a spherical gas cloud under the mutual attraction of its molecules and subject to the laws of classical thermodynamics.
The polytropic theory of stars essentially follows out of thermodynamic considerations, that deal with the issue of energy transport, through the transfer of material between different levels of the star.
We simply begin with the Poisson equation and the condition for hydrostatic equilibrium:
\begin{equation}
\frac{\mathrm{d}P}{\mathrm{d}r}=-\rho\frac{GM(r)}{r^2},
\end{equation}
\begin{equation}
\frac{\mathrm{d}M(r)}{\mathrm{d}r}=4\pi \rho{r^2},
\end{equation}
where $G$ is the gravitational constant, $P$ is the pressure at radius $r$, $M(r)$ is the mass of a star at a certain radius $r$, and $\rho$ is the density, at a distance $r$ from the center of a spherical star.
Combination of these equations yields the following equation, which as should be noted, is an equivalent form of the Poisson Equation.
\begin{equation}
\frac{1}{r^2}\frac{\mathrm{d}}{\mathrm{d}r}\left(\frac{r^2}{\rho}\frac{\mathrm{d}P}{\mathrm{d}r}\right)=-4\pi G\rho.
\end{equation}
From these equations one can obtain the Lane-Emden equation through the simple supposition that the pressure is related to the density, while remaining independent of the temperature.
We already know that in the case of a degenerate electron gas that the pressure and density are $\rho\thicksim P^\frac{3}{5}$, assuming that such a relation exists for other states of the star we are led to consider a relation of the following form:
\begin{equation}
P=K\rho^{1+\frac{1}{m}},
\end{equation}
where $K$ and $m$ are constants, at this point it is important to note that $m$ is the polytropic index which is related to the ratio of specific heats of the gas comprising the star.
Based upon these assumptions we can insert this relation into our first equation for the hydrostatic equilibrium condition and from this rewrite equation to:
\begin{equation}
\left[\frac{K(m+1)}{4\pi G}\lambda^{\frac{1}{m}-1}\right] \frac{1}{r^2}\frac{\mathrm{d}}{\mathrm{d}r}\left(r^2\frac{\mathrm{d}y}{\mathrm{d}r}\right)=-y^m,
\end{equation}
where the additional alteration to the expression for density has been inserted with $\lambda$ representing the central density of the star and $y$ that of a related dimensionless quantity that are both related to $\rho$ through the following relation
\begin{equation}
\rho=\lambda y^m.
\end{equation}
where $m$ is a constant.

Additionally, if place this result into the Poisson equation, we obtain a differential equation for the mass, with a dependance upon the polytropic index $m$.
Though the differential equation is seemingly difficult to solve, this problem can be partially alleviated by the introduction of an additional dimensionless variable $x$, given by the following:
\begin{equation}
r=ax,
\end{equation}
\begin{equation}
a=\left[\frac{K(m+1)}{4\pi G}\lambda^{\frac{1}{m}-1}\right]^\frac{1}{2}.
\end{equation}
Inserting these relations into our previous relations we obtain the famous form of the Lane-Emden equation, given below:
\begin{equation}
\frac{1}{x^2}\frac{\mathrm{d}}{\mathrm{d}x}\left(x^2\frac{\mathrm{d}y}{\mathrm{d}x}\right)=-y^m.
\end{equation}
Taking these simple relations we will have the Lane-Emden equation with $g(y)=y^m$,
\begin{equation}\label{Eq.Lane-Emden equation}
y''+\frac{2}{x}y'+y^m=0,\qquad x>0.
\end{equation}
The physically interesting range of m is $0 \leq m \leq 5$. Numerical and perturbation approaches to solve equation Eq. (\ref{Eq.Lane-EmdenMain}) with $g(y)=y^m$ and boundary conditions (\ref{Eq.Lane-EmdenBoundry}) have been considered by various authors. It has been claimed in the literature that only for $m$ = $0$, $1$ and $5$ the solutions of the Lane-Emden equation (also called the polytropic differential equations) could be given in closed form.

In fact, for $m = 5$, only a 1-parameter family of solutions is presented. The so-called generalized Lane-Emden equation of the first kind have been looked at in Goenner \cite{Goenner2001} and Havas \cite{Havas}.
\subsection{Methods have been used to solve Lane-Emden equations} 
Recently, many analytical methods have been used to solve Lane-Emden equations, the main difficulty arises in the
singularity of the equation at $x = 0$. Currently, most techniques in use for handling the Lane-Emden-type problems are based on either series solutions or perturbation techniques.\\
Bender et al. \cite{Bender.J. Math. Phys.1989}, proposed a new perturbation technique based on an
artificial parameter $\delta$, the method is often called $\delta$-method.\\
Mandelzweig et al. \cite{Mandelzweig} used Quasilinearization approach to solve Lane-Emden equation.
This method approximates the solution of a nonlinear differential equation by treating the nonlinear terms as a perturbation about the linear ones, and unlike perturbation theories is not based on the existence of some kind of a small parameter. He showed that the quasilinearization method gives excellent results when applied to different nonlinear ordinary differential equations in physics, such as the Blasius, Duffing, Lane-Emden and Thomas-Fermi equations.\\
Shawagfeh \cite{Shawagfeh} applied a nonperturbative approximate analytical solution for the Lane-Emden equation using the Adomian decomposition method. His solution was in the form of a power series. He used Pad\'{e} approximants method to accelerate the convergence of the power series.\\
In \cite{Wazwaz.Appl}, Wazwaz employed the Adomian decomposition method with an alternate framework designed to overcome the difficulty of the singular point.
It was applied to the differential equations of Lane-Emden type.
Further in \cite{Wazwaz.Appl.2006} he used the modified decomposition method for solving analytical treatment of nonlinear differential equations such as Lane-Emden equation.
The modified method accelerates the rapid convergence of the series solution, dramatically reduces the size of work, and provides the solution by using few iterations only without any need to the so-called Adomian polynomials.\\
Liao \cite{Liao.Appl. Math. Comput.2003} provided a reliable, easy-to-use analytical algorithm for Lane-Emden type equations. This algorithm logically contains the well-known Adomian decomposition method. Different from all other analytical techniques, this algorithm itself provides us with a convenient way to adjust convergence regions even without Pad\'{e} technique.\\
He \cite{He.AMC2003} employed Ritz's method to obtain an analytical solution of the problem. By the semi-inverse method, a variational principle is obtained for the Lane-Emden equation, which he gave much numerical convenience when applied to finite element
methods or Ritz method.\\
Parand et al. \cite{Parand.phy2004,Parand.mat2004,Parand.com2004} presented some numerical techniques to solve higher ordinary differential equations such as Lane-Emden.
Their approach was based on a rational Chebyshev and rational Legendre tau method.
They presented the derivative and product operational matrices of rational Chebyshev and rational Legendre functions.\\
These matrices together with the tau method were utilized to reduce the solution of these physical problems to the solution of systems of algebraic equations. Parand et al. \cite{Parand.JCP} also applied pseudospectral method based on rational Legendre functions to solve Lane-Emden equations.\\
Ramos \cite{Ramos.Appl. Math. Comput.2005,Ramos.Chaos Soliton. Frac.2007,Ramos.Chaos Soliton. Frac.2008,Ramos.Comp. Phys. Commmun.2003} solved Lane-Emden equation through different methods.
In \cite{Ramos.Appl. Math. Comput.2005} he presented linearization methods for singular initial-value problems in second-order ordinary differential equations such as Lane-Emden.
These methods result in linear constant-coefficients ordinary differential equations which can be integrated analytical, thus yielding piecewise analytical solutions and globally smooth solutions.
Later, he \cite{Ramos.Chaos Soliton. Frac.2007} developed piecewise-adaptive decomposition methods for the solution of nonlinear ordinary differential equations.
Piecewise-decomposition methods provide series solutions in intervals which are subject to continuity conditions at the end points of each interval, and their adaption is based on the use of either a fixed number of approximants and a variable step size, a variable number of approximants and a fixed step size or a variable number of approximants and a variable step size.\\
In \cite{Ramos.Chaos Soliton. Frac.2008}, series solutions of the Lane-Emden equation have been obtained by writing this equation as a Volterra integral
equation and assuming that the nonlinearities are sufficiently differentiable. These series solutions have been obtained by either working with the original differential equation or transforming it into an ordinary differential equation that does not contain first-order derivatives. It has been shown that these approaches provide exactly the same solutions as those based on Adomian's decomposition techniques that make use of either a different differential operator that overcomes the singularity at $x = 0$, or a new dependent variable, and Liao's homotopy analysis technique.
Series solutions to the Lane-Emden equation have also been obtained by working directly on the original differential equation or transforming it into a simpler one.\\
Yousefi \cite{Yousefi} presented a numerical method for solving the Lane-Emden equations as singular initial value problems. Using integral
operator and convert Lane-Emden equations to integral equations and then applying Legendre wavelet approximations. He presented Legendre wavelet properties and then utilized these properties together with the Gaussian integration method to reduce the integral equations to the solution of algebraic equations.\\
In \cite{Hashim}, Chowdhury et al. presented a reliable algorithm based on the homotopy-perturbation method (HPM) to solve singular IVPs of time-independent equations.they obtained the approximate and/or exact analytical solutions of the generalized Emden-Fowler type equations.This method is a coupling of the perturbation method and the homotopy method.The HPM is a novel and effective method which can solve various nonlinear equations.
The main feature of the HPM is that it deforms a difficult problem into a set of problems which are easier to solve.
In this work, HPM yields solutions in convergent series forms with easily computable terms.\\
Aslanov \cite{Aslanov.Phys. Lett. A2008} introduced a further development in the Adomian decomposition method to overcome the difficulty at the singular point of non-homogeneous, linear and non-linear Lane-Emden-like equations; and constructed a recurrence relation for the components of the approximate solution and investigated the convergence conditions for the Emden-Fowler type of equations. He improved the previous results on the convergence radius of the series solution.
Recently, Dehghan and Shakeri \cite{Dehghan.New Astron.2008} first applied an exponential transformation to the Lane-Emden equation to overcome the difficulty of a singular point at $x = 0$ and solved the resulting nonsingular problem by the variational iteration method. Yildirim et al. \cite{Yildirim} presented approximate exact solutions of a class of Lane-Emden type singular IVPs problems, by the variational iteration method. The variational iteration method yields solutions in the forms of convergent series with easily calculable terms.
Bataineh et al. \cite{Bataineh.Commun. Nonlinear Sci. Numer. Simul.2008} presented a reliable algorithm based on HAM to obtain the exact and/or approximate analytical solutions of the singular IVPs of the Emden-Fowler type. The HAM, first proposed by Liao in his Ph.D. dissertation \cite{Liao.Appl. Math. Comput.2003}, is a promising method for linear and non-linear problems. HAM contains an auxiliary parameter $\hbar$ which provides us with a simple way to adjust and control the convergence region and the rate of convergence of the series solution.\\
Marzban et al. \cite{Marzban} used a method based upon hybrid function approximations. He used the properties of hybrid of block-pulse functions and Lagrange interpolating polynomials together with the operational integration matrix for solving nonlinear second-order, initial value problems and the Lane-Emden equation.

\section{Properties of modified generalized Laguerre functions}\label{Properties.of.MGLF} 
This section is devoted to the introduction of the basic notions and working tools concerning orthogonal modified generalized Laguerre functions. More specifically, we presented some properties of modified generalized Laguerre functions, concerning projection process.

The Laguerre approximation has been widely used for numerical solutions of differential equations on infinite intervals.
$L_n^{\alpha}(x)$ (generalized Laguerre polynomial) is the $n$th eigenfunction of the Sturm-Liouville problem \cite{FunaroPaper,FunaroBook,SELCUK}:
\begin{eqnarray} \nonumber
x\frac{d^2}{dx^2}L_{n}^{\alpha}(x)+({\alpha}+1-x)\frac{d}{dx}L_{n}^{\alpha}(x)+nL_{n}^{\alpha}(x)=0,\\ \nonumber
x\in (0,\infty), \qquad n=0,1,2,....\quad,
\end{eqnarray}
with the normalizing condition:
\begin{eqnarray}\nonumber
&&L_n^{\alpha}(0)=\binom{n+\alpha}{n}\\ \nonumber
&&\frac{d}{dx}L_0^{\alpha}(x)=0, \quad \frac{d}{dx}L_1^{\alpha}(x)=-1,\nonumber
\end{eqnarray}
where $\alpha>-1$.\\
The generalized Laguerre polynomials can be defined with the following recurrence formula:
\begin{eqnarray}\nonumber
&&L_0^{\alpha}(x)=1, \quad L_1^{\alpha}(x)=1+\alpha-x, \\ \nonumber
&&nL_{n}^{\alpha}(x)=(2n-1+\alpha-x)L_{n-1}^{\alpha}(x)-(n+\alpha-1)L_{n-2}^{\alpha}(x), ~~{n\geq 2}
\end{eqnarray}
these are orthogonal polynomials for the weight function $w_\alpha=x^{\alpha}e^{-x}$.
The generalized Laguerre polynomials satisfy the following relation \cite{FunaroBook}:
\begin{eqnarray}\label{DerivateRelation}
\partial_xL_n^{\alpha}(x)=-\sum_{k=0}^{n-1}L_k^{\alpha}(x),
\end{eqnarray}
where $n\geq 1$ and $\alpha > -1$.\\
We define Modified generalized Laguerre functions (which we denote MGL functions) $\Gamma_{n}^{\alpha}(x)$ as follows:
\begin{eqnarray}\label{MGLF}
\Gamma_{n}^{\alpha}(x)=\exp({-x/(2L)})L_n^{\alpha}(x/L),\quad L > 0~~ \text{and}~~\alpha>-1.
\end{eqnarray}
This system is an orthogonal basis \cite{Boyd2000,George.Gasper} with weight function $w_{L}(x)=\frac{x}{L}$ and orthogonality property:
\begin{eqnarray}\nonumber
<\Gamma_{n}^{\alpha},\Gamma_{m}^{\alpha}>_{w_{L}}=\left(\frac{\Gamma(n+\alpha+1)}{L^2{n}!}\right)\delta_{nm},
\end{eqnarray}
where $\delta_{nm}$ is the Kronecker function.
Boyd \cite{Boyd2000,Boyd2003,Boyd.J. Comput. Phys.1982} offered guidelines for optimizing the map parameter $L$ where $L>0$ is the scaling parameter. On a semi-infinite domain, there is always a parameter that must be determined experimentally.\\
Numerical results deponed smoothly on constant parameter $L$, and therefore, are not very sensitive to $L$ because the ${d Error}/{dL}=0$ at the minimum itself, so the error varies very slowly with $L$ around the minimum. A little trial and error is usually sufficient to find a value that is nearly optimum. In general, there is no way to avoid a small amount of trial and error in choosing $L$ when solving problems on an unbounded domain. Experience and the asymptotic approximations of \cite{Boyd.J. Comput. Phys.1982} can help, but some experimentation is always necessary as he explain in his book \cite{Boyd2000}.\\

\section{Lagrangian interpolation}\label{Sec.Lagrangian.Interpolation} 
In this section we detailed description of Lagrangian interpolants construction and properties, then we introduce the Lagrangian interpolation of modified generalized Laguerre functions and its operational matrices of derivative is obtained.

Let $\{L_{k}^{\alpha}\}_{0\leq k\leq n}$ is a basis in the space $P_n$ of polynomials
of degree at most $n$. When $n$ distinct points are given, another basis in $P_n$
is generated in a natural way. This is the basis of Lagrange polynomials with respect
to the prescribed points. An element of the basis attains the value $1$ at a certain point
and vanishes in the remaining $n-1$ points.

Let us analyze first the generalize Laguerre polynomial case. We have the set of the Lagrange polynomials in $P_n$ relative to the $n$ points $\eta_{k}$, $0\leq k \leq n-1$, i.e., the zeroes of $L_{n}^{\alpha}$.

The elements of the basis are denoted by $\ell_{j}^{n}~~0 \leq j\leq n-1$ (denoted Lagrangian interpolants). These
polynomials in $P_n$ are uniquely defined by the conditions
\begin{equation}
\ell_{j}^{n}(\eta_{i}) =
\begin{cases}
1&\quad \text{if}~~ i=j\\
0&\quad \text{if}~~ i\neq j
\end{cases}
,~~0 \leq j \leq n-1.
\end{equation}
They actually form a basis because any polynomial $p \in P_{n}$ can be written as follows:
\begin{eqnarray}\nonumber
p=\sum_{j=0}^{n-1}p(\eta_j)\ell_{j}^{n}.
\end{eqnarray}
Therefore, $p$ is a linear combination of the Lagrange polynomials. Such a combination
is uniquely determined by the coefficients $p(\eta_j),~~ 0\leq j \leq n-1$.\\
The following expression is easily proven:
\begin{eqnarray}\nonumber
\ell_{j}^{n}(x)=\prod_{k=0,k\neq j}^{n-1}\frac{x-\eta_k}{\eta_j-\eta_k},~~0 \leq j \leq n-1.
\end{eqnarray}
For future applications, it is more convenient to consider the alternate expression
\begin{equation}
\ell_{j}^{n}(x) =
\begin{cases}
\frac{L_{n}^{\alpha}(x)}{{L'}_{n}^{\alpha}(\eta_j)(x-{\eta_j})}& \quad \text{if}~~ x\neq \eta_j,\\
1&\quad \text{if}~~ x=\eta_j,
\end{cases}
\end{equation}
where $0 \leq j \leq n-1$. Of course, we have
\begin{eqnarray}\nonumber
\lim_{x\rightarrow\eta_j}\ell_{j}^{n}(x)=\lim_{x\rightarrow\eta_j}\frac{{L'}_{n}^{\alpha}(x)}{{L'}_{n}^{\alpha}(\eta_j)}=1,~~0 \leq j \leq n-1.
\end{eqnarray}

Lagrangian interpolants of generalized Laguerre polynomials (we denoted GLP) of order $p$ at the Gauss-Radau-Laguerre quadrature points in $\mathbb{R}^{+}$ is \cite{FunaroBook}:
\begin{equation}\label{FistLagrangian}
\ell_{j}^{n}(x) =
\begin{cases}
\frac{xL_{n}^{\alpha}(x)}{\eta_j{L'}_{n}^{\alpha}(\eta_j)} \frac{1}{x-\eta_j},&\qquad j=1,...,n,\\
\\
\frac{L_{n}^{\alpha}(x)}{L_{n}^{\alpha}(0)},&\qquad j=0,
\end{cases}
\end{equation}
where
$\eta_j ,j=0,1,2,...,n$ are the $n+1$ GLP-Radau points.\\
derivative operator of GLP is:
\begin{eqnarray}\nonumber
d_{ij} =\ell^{'n}_j(\eta_i),
\end{eqnarray}
moreover for any polynomial $p$ of degree at most $n+1$, one gets:
\begin{eqnarray}\nonumber
p^{'}(\eta_i)=\sum_{j=0}^{n}{d_{ij}p(\eta_j)}.
\end{eqnarray}
Funaro \cite{FunaroPaper,FunaroBook} obtained derivative matrix of GLP($D_n$):

\begin{equation}\label{FirstDerivateMatrix}
d_{ij}=
\begin{cases}
\frac{\eta_i\frac{d}{dx}L_{n}^{\alpha}(\eta_i)}{\eta_j\frac{d}{dx}L_{n}^{\alpha}(\eta_j)}\frac{1}{\eta_i-\eta_j} & i,j=1,...,n , i\neq j,\\
\frac{1-\alpha+\eta_i}{2\eta_i} & i=j=1,...,n,\\
\frac{\frac{d}{dx}L_{n}^{\alpha}(\eta_i)}{L_{n}^{\alpha}(0)} & i=1,...,n ,j=0,\\
-\frac{L_{n}^{\alpha}(0)}{\eta_j^{2}\frac{d}{dx}L_{n}^{\alpha}(\eta_j)} & j=1,...,n ,i=0,\\
-\frac{n}{\alpha+1} & i=j=0.\\
\end{cases}
\end{equation}

The second derivative operator is obtained either by squaring $D_n$ either by evaluating $\ell^{''n}_{ij}(\eta_i)$:
\begin{equation}\label{SecondDerivateMatrix}
\ell^{''n}_{ij}(\eta_i) =
\begin{cases}
\frac{\frac{d}{dx}L_{n}^{\alpha}(\eta_i)((1-\alpha+\eta_i)(\eta_i-\eta_j)-2\eta_i)}{\eta_j(\eta_i-\eta_j)^{2}\frac{d}{dx}L_{n}^{\alpha}(\eta_j)} & i,j=1,...,n , i\neq j,\\
\frac{(\eta_i-\alpha)^2}{3\eta_i^2}-\frac{n-1}{3\eta_i} & i=j=1,...,n, \\
-\frac{(\alpha+1-\eta_i)\frac{d}{dx}{L_{n}^{\alpha}(\eta_i)}}{\eta_i{L_{n}^{\alpha}(0)}} & i=1,...,n ,j=0, \\
-\frac{{2(n+\alpha+1}){L_{n}^{\alpha}(0)}}{{{\eta_{j}^{3}(\alpha+1)}\frac{d}{dx}L_{n}^{\alpha}(\eta_j)}} & j=1,...,n ,i=0, \\
\frac{n(n-1)}{(\alpha+1)(\alpha+2)} & i=j=0. \\
\end{cases}
\end{equation}
Laguerre polynomials are not suitable for computations \cite{FunaroBook}, and also
Lagrangian interpolation of Laguerre polynomials is not suitable for solving some differential equations, such as Lane-Emden equations because of their boundary conditions. So we use Lagrangian interpolation of MGL functions. At first we must find Lagrangian interpolants and derivative operators of MGL functions.

Let $\Gamma_{n}^{\alpha}(x)=e^{-x/2}L_{n}^{\alpha}(x)$ and by substitution of $\eta_i$ with $x$ we have,
\begin{equation}\label{DerriveOFGamma}
\frac{d}{dx}\Gamma_{n}^{\alpha}(x)\Big|_{x=\eta_i} =e^{-x/2}\frac{d}{dx}L_{n}^{\alpha}(x)\Big|_{x=\eta_i}.
\end{equation}

\begin{lemma}
Lagrangian interpolant of $\Gamma_{n}^{\alpha}(x)=e^{-x/2}L_{n}^{\alpha}(x)$ is,
\begin{equation}\label{InterpolantFunc}
\widehat{\ell}_{i}^{n}(x) = \ell_{i}^{n}(x)\frac{e^{-x/2}}{e^{-\eta_i /2}},
\end{equation}
\end{lemma}
which $\ell_{j}^{n}(x)$ are Lagrangian interpolants of Laguerre polynomials.

Proof: Suppose $\gamma\frac{x\Gamma_{n}^{\alpha}(x)}{x-\eta_j}$ is
Lagrangian interpolant of $\Gamma_{n}^{\alpha}(x)$, using relation (\ref{DerriveOFGamma}) and L'H\^{o}pital's rule we can find constant
$\gamma$,
\begin{eqnarray}\nonumber
&& \widehat{\ell}_{j}^{n}(\eta_i)=0, \quad i\neq j\\ \nonumber
&& \lim_{x\rightarrow\eta_j}\gamma\frac{x\Gamma_{n}^{\alpha}(x)}{x-\eta_j}
=\gamma\lim_{x\rightarrow\eta_j} (\Gamma_{n}^{\alpha}(x)+x\frac{d}{dx}\Gamma_{n}^{\alpha}(x))=1\\ \nonumber &&\Rightarrow\gamma=\frac{1}{\eta_je^{-\eta_j /2}\frac{d}{dx}L_{n}^{\alpha}(x)\Big|_{x=\eta_j}},
\end{eqnarray}
so
\begin{equation}\label{SecondLagrangian}
\widehat{\ell}_{j}^{n}(x) = \frac{1}{\eta_je^{-\eta_j /2}\frac{d}{dx}L_{n}^{\alpha}(x)\Big|_{x=\eta_j}}\frac{xe^{-x/2}{L}_{n}^{\alpha}(x)}{x-\eta_j},
\end{equation}
and with comparison of (\ref{FistLagrangian}) and (\ref{SecondLagrangian}) Lemma 1 is proved.\\
By Eq. (\ref{InterpolantFunc}) derivative operator of $\Gamma_{n}^{\alpha}(x)=e^{-x/2}L_{n}^{\alpha}(x)$ (we denote by $\widehat{D}_n$) is:
\begin{equation}
\widehat{d}_{ij} = d_{ij}\frac{e^{-\eta_j/2}}{e^{-\eta_i/2}}-1/2\delta_{ij},\qquad i,j=0,...,n.
\end{equation}
Where matrix $d_{ij}$ is defined in Eq. (\ref{FirstDerivateMatrix}).
As pointed out in before the second derivative operator is obtained either by squaring $\widehat{D}_n$ either by evaluating $\widehat{\ell}^{''n}_j(\eta_i)$.
For evaluating $\widehat{\ell}^{''n}_j(\eta_i)$ we can use the following relation:
\begin{equation}
\widehat{\ell^{''n}_{ij}}(\eta_j)=\frac{1}{4}\delta_{ij}-d_{ij}\frac{e^{-\eta_j/2}}{e^{-\eta_i/2}}+\frac{e^{-\eta_j/2}}{e^{-\eta_i/2}}\ell^{''n}_{ij}(\eta_j),\qquad i,j=0,...,n,
\end{equation}
and $\ell^{''n}_{ij}(\eta_j)$ is defined in Eq. (\ref{SecondDerivateMatrix}).
It is obvious that MGL functions is $\Gamma_{n}^{\alpha}(x/L)$, so Lagrangian interpolant of MGL functions is $\widehat{\ell}_{i}^{n}(x/L)$, and derivative operators can be obtained easily.

\subsection{Function Approximation}
We define the interpolant approximation of $y(x)$ by
\begin{equation}\label{InfinityExpandU2}
I_{n} y(x)=\sum_{j=0}^{n}b_j\widehat{\ell}^{n}_{j}(x/L).
\end{equation}
Where $\widehat{\ell}^{n}_{j}(x)$ is defined in Eq. (\ref{SecondLagrangian}).
The $b_j$'s are the expansion coefficients associated with the family $\{\widehat{\ell}_{j}^{n}(x/L)\}$.
A semilogarithmic plot of $|b_j|$ versus $j$ is also useful to determine a good choice of $L$ when the exact solution for $y(x)$ is unknown. One can run the code for several different $L$ and then plot the coefficients from each run on the same graph. The best $L$ is the choice that gives the most rapid decrease of the coefficients \cite{Boyd2000}.\\
Therefore,
\begin{equation}\label{FindCoeficeint}
I_{n} y(\Im_j)=b_j,\qquad j=0,...,n.
\end{equation}
where $\Im_j=L\eta_j$ are the zeroes of $\widehat{\ell^{n}}_{j}(x/L)$.
So derivative operator of MGL functions is (we denote by $\widehat{D_L}_n$):
\begin{equation}
\widehat{d_{L_{ij}}}\label{Approximatfirst} =\frac{1}{L}\widehat{\ell^{'n}_j}(\Im_i/L)=\frac{1}{L}\widehat{\ell^{'n}_j}(\eta_i)=\frac{1}{L}\widehat{d_{ij}},
\end{equation}
and the second derivative operator is
\begin{equation}
\widehat{d^{(2)}_{L_{ij}}} =\frac{1}{L^2}\widehat{\ell^{''n}_j}(\Im_i/L)=\frac{1}{L^2}\widehat{\ell^{''n}_j}(\eta_i),
\end{equation}
The relationship between the derivative $\frac{d}{dx}I_{n} y(x)$ and $I_{n} y(x)$ at the collocation points $\Im_i,\quad i=0,...,n$ can be obtained by differentiation. The result is as:
\begin{equation}
\frac{d}{dx}I_{n} y(x)\Big|_{x=\Im_i}=\sum_{j=0}^{n}b_j\widehat{d_{L_{ij}}}\Big|_{x=\Im_i},~~ 0\leq i\leq n,
\end{equation}
and
\begin{equation}\label{ApproximatLast}
\frac{d}{d^{2}x}I^{2}_{n} y(x)\Big|_{x=\Im_i}=\sum_{j=0}^{n}b_j\widehat{d^{(2)}_{L_{ij}}}\Big|_{x=\Im_i},~~ 0\leq i\leq n.
\end{equation}

\section{Solving Lane-Emden equation}\label{solving.Lane-Emden} 
To apply Lagrangian interpolant of MGL functions to the Lane-Emden Equation introduced in Eq. (\ref{Eq.Lane-EmdenMain}) and Eq. (\ref{Eq.Lane-Emden}) with boundary conditions Eq. (\ref{Eq.Lane-EmdenBoundry}), at first with Eq. (\ref{InfinityExpandU2}) we expand $y(x)$, as follows:
\begin{eqnarray} \nonumber
I_{n} y(x)=\sum_{j=0}^{n}b_j\widehat{\ell^{n}_{j}}(x/L),
\end{eqnarray}
with $\alpha=1$ for Lagrangian interpolant of MGL functions.

To find the unknown coefficients $b_j$'s, we substitute the truncated series into the Eq. (\ref{Eq.Lane-EmdenMain}) with $g(y)$ introduced in Eq. (\ref{Eq.Lane-Emden}) and boundary conditions in Eq.~(\ref{Eq.Lane-EmdenBoundry}). So we have:
\begin{equation}\label{ChangedT.Lane-emden.}
\Im_i\sum_{j=0}^{n}b_j\widehat{d^{(2)}_{L_{ij}}}+2\sum_{j=0}^{n}b_j\widehat{d_{L_{ij}}}+\Im_ib_i^{m}=0 ,\quad i=1,...,n-1
\end{equation}
\begin{eqnarray}\label{ChangedT.Lane.Bound.}
&&b_0=1,\quad\\ \nonumber
&&\displaystyle\sum_{j=0}^{n}b_j\widehat{d_{L_{0j}}}=0.
\end{eqnarray}
We have $n-1$ equations in Eq. (\ref{ChangedT.Lane-emden.}), that generates a set of $n+1$ nonlinear equations with boundary equations in Eq. (\ref{ChangedT.Lane.Bound.}).

Table \ref{tbl1} shows the approximations of $y(x)$ for standard Lane-Emden with $m=3$ obtained by the method proposed in this paper for $n=7$ and $L=1$, and those obtained by Horedt \cite{Horedt}.

Table \ref{Tbl2} shows the comparison of the first zero of $y$, between the present method, rational Legendre pseudospectral method \cite{Parand.JCP}, method in \cite{Marzban}, Pad\'{e} approximation used by \cite{Bender.J. Math. Phys.1989} and exact values reported in \cite{Horedt} for $m = 2,3,4$, respectively.

\begin{figure}
\includegraphics[scale=0.5]{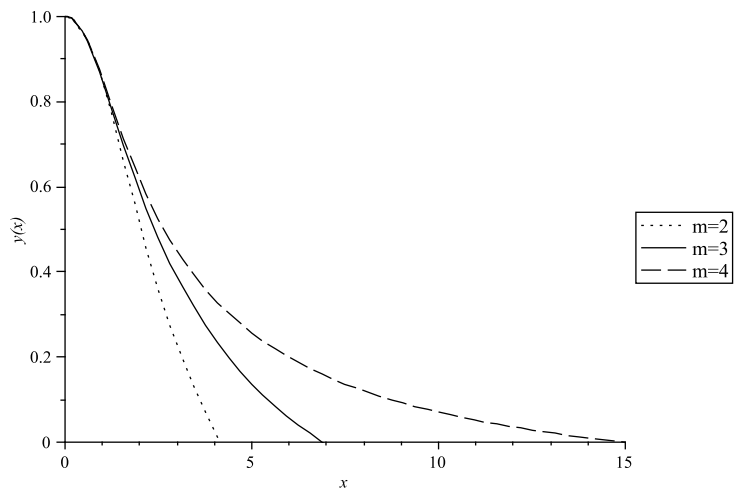}
\caption{Lane-Emden equation graph obtained by present method}
\label{fig}
\end{figure}

Figure \ref{fig} shows the result graph of Lane-Emden for $n=6$, $m=2,3,4$.

\section{Conclusions}\label{conclusion} 
The Lane-Emden equation describes a variety of phenomena in theoretical physics and astrophysics,
including aspects of stellar structure,
the thermal history of a spherical cloud of gas, isothermal gas spheres, and thermionic currents \cite{Chandrasekhar.Dover1957}.\\
Since then the equation has been a center of attraction.
The main problem in this direction is the accuracy and range of applicability of these approaches. The fundamental goal of this paper has been to construct an approximation to the solution of nonlinear Lane-Emden equation in
a semi-infinite interval which has singularity at $x=0$. A set of Laguerre functions are proposed to provide an effective but simple way to improve the convergence of the solution by Lagrangian method.
Through the comparisons among the exact solutions of Horedt \cite{Horedt}, the approximate
solutions of Bender \cite{Bender.J. Math. Phys.1989}, recent good results in \cite{Parand.JCP} and the current work, it has been shown that the present work has provided more accurate solutions for Lane-Emden equations. by taking to account $N$, it seems that the present approach gives the good result with small $N$.

\section*{Acknowledgments}
The authors are very grateful to the reviewers for carefully reading the Paper and for his (or her) comments and suggestions which have improved the Paper. One of the authors (K. Parand) would like to thank Shahid Beheshti University for the awarded grant.

\clearpage

\begin{table}
\caption{Approximation of $y(x)$ for present method, solutions of Horedt \cite{Horedt} for $m=3$}
\begin{tabular*}{\columnwidth}{@{\extracolsep{\fill}}*{3}{c}}
\hline
$x$ & Present method & solutions of Horedt \cite{Horedt}\\
\hline
$0.000$ & $1.000000$ & $1.000000$  \\
$0.100$ & $0.998323$ & $0.998336$  \\
$0.500$ & $0.959821$ & $0.959839$  \\
$1.000$ & $0.855057$ & $0.855058$  \\
$5.000$ & $0.110820$ & $0.110820$  \\
$6.000$ & $0.043718$ & $0.043738$  \\
$6.800$ & $0.004165$ & $0.004168$  \\
$6.896$ & $0.000035$ & $0.000036$  \\
\hline
\end{tabular*}\label{tbl1}\end{table}
\begin{table}
\caption{Comparison the first zero of $y$, between the present method, rational Legendre pseudospectral method \cite{Parand.JCP}, method in \cite{Marzban}, Pad\'{e} approximation used by \cite{Bender.J. Math. Phys.1989} and exact values reported in \cite{Horedt} for $m = 2,3$ and $4$}
\begin{tabular*}{\columnwidth}{@{\extracolsep{\fill}}*{7}{c}}
\hline
    &     &  Present        &   method in \cite{Parand.JCP} &  method    &          &                \\
$m$ & $N$ &  method         &   with $N=75$         &  in \cite{Marzban} & Bender& Exact value       \\
\hline
$2$ & $6$ &  $4.352875$     &   $4.35287108$        & $4.352875$         & $4.3603$  & $4.35287460$  \\
$3$ & $7$ &  $6.896849$     &   $6.89684862$        & $6.89685$          & $7.0521$  & $6.89684862$  \\
$4$ & $6$ &  $14.971546$    &   $14.9715463$        & $14.971546$        & $17.967$  & $14.9715463$  \\
\hline
\end{tabular*}
\label{Tbl2}
\end{table}
\clearpage

\end{document}